\documentstyle[12pt]{article}

\newcounter{eq}
\newcounter{sc}

\def\overleftrightarrow#1{\vbox{\ialign{##\crcr
 $\leftrightarrow$\crcr\noalign{\kern-1pt\nointerlineskip}
 $\hfil\displaystyle{#1}\hfil$\crcr}}}

\newlength{\minitwocolumn}
\begin{document}

\begin{flushright}
DPUR/TH/58\\
November, 2017\\
\end{flushright}
\vspace{20pt}

\pagestyle{empty}
\baselineskip15pt

\begin{center}
{\large\bf Cosmic Acceleration in the Nonlocal Approach to the Cosmological Constant Problem
\vskip 1mm }

\vspace{10mm}
Ichiro Oda\footnote{E-mail address:\ ioda@phys.u-ryukyu.ac.jp
}

\vspace{3mm}
           Department of Physics, Faculty of Science, University of the 
           Ryukyus,\\
           Nishihara, Okinawa 903-0213, Japan.\\

\end{center}


\vspace{3mm}
\begin{abstract}
We have recently constructed a manifestly local formulation of a nonlocal approach to the cosmological constant problem 
which can treat with quantum effects from both matter and gravitational fields. In this formulation, it has been explicitly 
shown that the effective cosmological constant is radiatively stable even in the presence of the gravitational loop effects. 
Since we are naturally led to add the $R^2$ term and the corresponding topological action to an original
action, we make use of this formulation to account for the late-time acceleration of expansion of the universe in case of 
the open universes with infinite space-time volume. We will see that when the "scalaron", which exists in the $R^2$ gravity
as an extra scalar field, has a tiny mass of the order of magnitude ${\cal{O}}(1 meV)$, we can explain the current value of the
cosmological constant in a consistent manner.  
\end{abstract}

\newpage
\pagestyle{plain}
\pagenumbering{arabic}


\rm
\section{Introduction}

The recent development of observational cosmology has made it possible to tell that our universe might have 
undergone two phases of cosmic acceleration \cite{Riess}-\cite{Ade}.  The first stage of the acceleration is 
usually called $\it{inflation}$ and is believed to have happened prior to the radiation-dominated era 
\cite{Guth}-\cite{Albrecht}. The inflation is needed not only to resolve the flatness and horizon problems 
but also to account for the almost flat spectrum 
of temperature anisotropies observed in the cosmic microwave background. An attractive feature 
of the inflation is that it provides a mechanism to generate the small scale fluctuations which are required 
to seed galaxy formation. These fluctuations are nothing but quantum fluctuations due to the zero-point energy 
of a generic quantum field, which get pushed into the classical regime by the large expansion of the
universe. This accelerated expansion must end to connect to the radiation-dominated universe via a period
of reheating, so the pure cosmological constant should not be responsible for inflation. 

The fact that we live in a large and old universe with a large amount of entropy can also be regarded as a primary
consequence of inflation in the early universe. The standard paradigm of inflation is that the accelerated expansion
of the universe is realized when a scalar field, which is usually dubbed as the $\it{inflaton}$, slowly rolls over its
potential down to a global minimum, which is identified with the true vacuum, in a time scale longer than 
the Hubble time. However, the origin of inflaton and its potential is quite vague, which is one important issue to 
be clarified in future.

Actually, the slow-roll inflation is not the only inflation scenario studied so far, and there are a few alternative theories. 
Among them, what we call, the Starobinsky model of inflation does not require introduction of the inflaton by hand 
and the scalar field degree of freedom, which is specially called "scalaron", emerges from the higher-derivative
curvature term $R^2$ with a gravitational origin \cite{Staro}. Unlike the inflationary models such as "old inflation", 
the Starobinsky model is not plagued by the graceful exit problem since the accelerated expansion of the universe is 
triggered by the $R^2$ term whereas the followed radiation-dominated epoch with a transient matter-dominated phase
is caused by the Einstein-Hilbert term.

On the other hand, in order to explain the second stage of the acceleration, it might be possible to make use of 
the cosmological constant since the acceleration in "recent" years does not have to end and could last forever.
However, at this point we encounter one of the most serious contradictions in modern physics, the cosmological constant
problem, which implies the enormous mismatch between the observed value of cosmological constant and estimate of 
the contributions of elementary particles to the vacuum energy density \cite{Weinberg, Padilla}.
  
For later convenience, within the context of quantum field theory (QFT), let us explain the cosmological 
constant problem by using a real scalar field of mass $m$ with $\lambda \phi^4$ interaction, which is 
minimally coupled to the $\it{classical}$ gravity \cite{Oda7}\footnote{We follow notation and conventions of the textbook 
by Misner et al \cite{MTW}.} 
\begin{eqnarray}
S  = \int d^4 x \sqrt{- g} \ \Biggl[ \frac{M_{Pl}^2}{2} R - \Lambda_b - \frac{1}{2} g^{\mu\nu} \partial_\mu \phi \partial_\nu \phi
- \frac{m^2}{2} \phi^2 - \frac{\lambda}{4!} \phi^4  \Biggr],
\label{phi4}
\end{eqnarray}
where $M_{Pl}$ is the reduced Planck mass defined as $M_{Pl} = \frac{1}{\sqrt{8 \pi G}}$ ($G$ is the Newton constant), 
and $\Lambda_b$ is the bare cosmological constant which is in principle divergent. Using the dimensional regularization, 
the 1-loop effective potential can be calculated to be
\begin{eqnarray}
V^{\phi, 1-loop} &\equiv& \frac{i}{2} tr \left[ \log \left( - i \frac{\delta^2 S}{\delta \phi^2} \right) \right]
= \frac{1}{2} \int \frac{d^4 k_E}{(2 \pi)^4} \log ( k_E^2 + m^2 )
\nonumber\\
&=&  - \frac{m^4}{(8 \pi)^2} \left[ \frac{2}{\epsilon}  + \log \left( \frac{\mu^2}{m^2} \right) + finite \right],
\label{V-1 loop}
\end{eqnarray}
where $\mu$ is the renormalization mass scale. In order to cancel the divergence associated with a simple pole $\frac{2}{\epsilon}$,
we are required to choose the bare cosmological constant at the 1-loop level
\begin{eqnarray}
\Lambda^{\phi, 1-loop}_b = \frac{m^4}{(8 \pi)^2} \left[ \frac{2}{\epsilon}  + \log \left( \frac{\mu^2}{M^2} \right) \right],
\label{Bare-Lambda-1 loop}
\end{eqnarray}
where $M$ is an arbitrary subtraction mass scale where the measurement is carried out.  (In cosmology, the physical meaning 
of $M$ is not so clear as in particle physics since $M$ may be identified with the Hubble parameter or any other cosmological
energy scale related to the evolution of the universe \cite{Brax}.) Then, by summing up the two contributions,
the 1-loop renormalized cosmological constant is of form 
\begin{eqnarray}
\Lambda^{\phi, 1-loop}_{ren} = V^{\phi, 1-loop} + \Lambda^{\phi, 1-loop}_b
= \frac{m^4}{(8 \pi)^2} \left[ \log \left( \frac{m^2}{M^2} \right) - finite \right].
\label{Renorm-Lambda-1 loop}
\end{eqnarray}

The cosmological observation requires us to take $\Lambda^{\phi, 1-loop}_{ren} \sim (1 meV)^4$. 
If the particle mass $m$ is chosen to the electro-weak scale, we have $V^{\phi, 1-loop} \sim (1 TeV)^4 = 10^{60} (1 meV)^4$. 
Thus, the measurement suggests that the finite contribution to the 1-loop renormalized cosmological constant is cancelled to an accuracy 
of one part in $10^{60}$ between $V^{\phi, 1-loop}$ and $\Lambda^{\phi, 1-loop}_b$. This big fine tuning is sometimes called the cosmological 
constant problem.\footnote{There is a loophole in this argument. If the particle mass $m$ is taken to be $1 meV$ like neutrinos,
we have $V^{\phi, 1-loop} \sim (1 meV)^4$, which is equivalent to the observed value of the cosmological constant, so in this case
there is no cosmological constant problem. Actually, we will meet such a situation later in discussing the scalaron mass.} Following the lore of QFT, 
at this stage of the argument, we have no issue with this fine tuning since we have just replaced the renormalized quantity depending
on the arbitrary scale $M$ with the measured value.

However, the issue arises when we further consider higher loops. For instance, at the 2-loop level, $V^{\phi, 2-loop}$ is proportional to 
$\lambda m^4$ (In general, at the n-loop level,  $V^{\phi, n-loop} \propto \lambda^{n-1} m^4$). Then, the consistency between 
the measurement and the perturbation theory requires us to set up an equality
\begin{eqnarray}
(1 meV)^4 = \Lambda_{ren} = \Lambda_b + V^{\phi, 1-loop} + V^{\phi, 2-loop} + V^{\phi, 3-loop} + \cdots.
\label{Renorm-Lambda-n loop}
\end{eqnarray}
The problem is that each $V^{\phi, i-loop} (i = 1, 2, \cdots)$ has almost the same and huge size compared to the observed value
$(1 meV)^4$.  Thus, even if we fined tune the cosmological constant at the 1-loop level, the fine tuning would be spoilt at the 2-loop level
so that we must retune the finite contribution in the bare cosmological constant term to the same degree of accuracy. In other words, 
at each successive order in perturbation theory, we are required to fine tune to extreme accuracy!  This problem is called "radiative instability", 
i.e., the need to repeatedly fine tune whenever the higher loop corrections are included, which is the essence of the cosmological constant
problem. 

In order to solve the problem of the radiative instability of the cosmological constant, some nonlocal formulations have been advocated
\cite{Linde}-\cite{Oda3}, but many of them have been restricted to the semiclassical approach where only radiative corrections 
from matter fields are taken account of whereas the gravity is regarded as a classical field merely serving for detecting the vacuum energy. 
Recently, an interesting approach, which attempts to deal with the graviton loop effects by using the topological Gauss-Bonnet term, 
has been proposed \cite{Kaloper3}.  

More recently, we have proposed an alternative formulation for treating with the graviton loops where the higher derivative term $R^2$ 
plays an important role \cite{Oda7}. In the article \cite{Oda7}, we have also pointed out a connection between the formulation and 
the $R^2$ inflation by Starobinsky \cite{Staro}. Thus far, there are a few articles dealing with inflation in the closed universes
with finite space-time volume by a similar formalism \cite{Kaloper1, Avelino}, but these papers are written in the framework of 
the semiclassical approach, so there is indeed radiative instability in the gravity sector. In this article, we would like to apply 
our formulation \cite{Oda7}, which is completely free from the issue of radiative instability, to the late-time acceleration of expansion 
of the universe in the open universes with infinite space-time volume.  The reasons why we consider the open universes are two-fold. 
Current cosmological data show as a whole good agreement with a spatially flat universe with non-vanishing cosmological constant and 
cold dark matter, which suggests that our universe might be open. Moreover, the scenario of eternal inflation leads to space-times 
with infinite volume \cite{Linde2}.

\section{Gravitational field equation with a high-pass filter}

It is an interesting direction of research to explore a possibility of constructing an effective theory with nonlocal properties
in order to address the cosmological constant problem. The fundamental physical idea that we wish to implement in the present study 
is to render the effective Newton constant depend on the frequency and wave length such that it works as a filter shutting off 
oscillating modes with high energy. Since we want to keep the fundamental physical principles such as locality and unitarity, 
we will attempt to simply modify general relativity at the level of not the action but the equations of motion.  
Because of it, we can present a solution to the cosmological constant problem without losing all the successes of general relativity. 
As seen later in the present article, this idea is only effective for making the vacuum energy density stemming from matter fields be 
radiatively stable, so we need an additional mechanism for shutting off the high energy modes of gravitational field.  
We will present such an additional mechanism by incorporating the $R^2$ term and the corresponding topological term
in the next section.

In Ref. \cite{Nima}, such a filter mechanism has been already proposed so we will make use of this mechanism, but the obtained 
equation in the present article is different from that in Ref.  \cite{Nima} since we proceed along a slightly different path of arguments.
The filter mechanism advocated in \cite{Nima} is to start with the following modified Einstein equation:
\begin{eqnarray}
M_{Pl}^2 \left[ 1 + {\cal{F}} (L^2 \Box) \right]  G_{\mu\nu} = T_{\mu\nu},
\label{Filter-Eins-Eq1}
\end{eqnarray}
where $G_{\mu\nu} = R_{\mu\nu} - \frac{1}{2} g_{\mu\nu} R$ is the well-known Einstein tensor, $T_{\mu\nu}$ is the energy-momentum
tensor, and the filter function ${\cal{F}} (L^2 \Box)$ is assumed to take the form; ${\cal{F}} (x) \rightarrow 0$ for $x \gg 1$ 
whereas ${\cal{F}} (x) \gg 1$ for $x \ll 1$. Here $L$ is the length scale where gravity is modified, and $\Box \equiv
g^{\mu\nu} \nabla_\mu \nabla_\nu$ is a covariant d'Alambertian operator. It is a natural interpretation that we regard 
the above Einstein equation (\ref{Filter-Eins-Eq1}) as the Einstein equation with the effective Newton constant $\tilde G_{eff}$ 
or the effective reduced Planck mass $\tilde M_{eff}$ defined as 
\begin{eqnarray}
\tilde M_{eff}^2 \equiv \frac{1}{8 \pi \tilde G_{eff}} \equiv M_{Pl}^2 \left[ 1 + {\cal{F}} (L^2 \Box) \right].
\label{Effective-Newton}
\end{eqnarray}

As considered in Ref. \cite{Nima}, let us take the infinite length limit $L \rightarrow \infty$ to pick up the "zero mode" of $G_{\mu\nu}$
such that $G_{\mu\nu} = g_{\mu\nu}$. In this limit we obtain
\begin{eqnarray}
{\cal{F}} (L^2 \Box) G_{\mu\nu} = {\cal{F}} (0) g_{\mu\nu} \equiv c g_{\mu\nu},
\label{Zero-mode}
\end{eqnarray}
where we have defined the constant $c$ by $c = {\cal{F}} (0)$. Then, substituting Eq. (\ref{Zero-mode}) into Eq. (\ref{Filter-Eins-Eq1})
leads to the equation
\begin{eqnarray}
M_{Pl}^2 G_{\mu\nu} + c g_{\mu\nu} = T_{\mu\nu}.
\label{Filter-Eins-Eq2}
\end{eqnarray}
Taking the trace of this equation, we can describe the constant $c$ as
\begin{eqnarray}
c = \frac{1}{4} ( M_{Pl}^2 R + T ).
\label{c-rel1}
\end{eqnarray}
In general, the right-hand side (RHS) is not a constant, so to make the RHS be a constant as well, we will take the space-time average 
of the both sides:
\begin{eqnarray}
c = \frac{1}{4} ( M_{Pl}^2 \overline{R} + \overline{T} ).
\label{c-rel2}
\end{eqnarray}
Here, for a generic space-time dependent quantity $Q(x)$, the operation of taking the space-time average is defined as 
\begin{eqnarray}
\overline{Q(x)}  = \frac{\int d^4 x \sqrt{-g} \, Q(x)}{\int d^4 x \sqrt{-g}},
\label{ST average}
\end{eqnarray}
where the denominator $V \equiv \int d^4 x \sqrt{-g}$ denotes the space-time volume.  Finally, inserting Eq.  (\ref{c-rel2})
to Eq. (\ref{Filter-Eins-Eq2}), we arrive at the desired equation of motion for the gravitational field:
\begin{eqnarray}
M_{Pl}^2 G_{\mu\nu} + \frac{1}{4} M_{Pl}^2 \overline{R} g_{\mu\nu} = T_{\mu\nu} - \frac{1}{4} \overline{T} g_{\mu\nu}.
\label{Filter-Eins-Eq3}
\end{eqnarray}

It is easy to show that this new type of Einstein equation has a close relationship with unimodular gravity \cite{Einstein}. To do so,
let us first take the trace of Eq. (\ref{Filter-Eins-Eq3}): 
\begin{eqnarray}
M_{Pl}^2 \overline{R} + \overline{T} = M_{Pl}^2 R + T.
\label{Filter-Eins-Eq3-Tr}
\end{eqnarray}
Substituting Eq. (\ref{Filter-Eins-Eq3-Tr}) into Eq.  (\ref{Filter-Eins-Eq3}) gives rise to the equation of motion in unimodular
gravity
\begin{eqnarray}
R_{\mu\nu} - \frac{1}{4} g_{\mu\nu} R = \frac{1}{M_{Pl}^2} ( T_{\mu\nu} - \frac{1}{4} g_{\mu\nu} T).
\label{UG}
\end{eqnarray}
Actually, operating the covariant derivative $\nabla^\mu$ on Eq. (\ref{UG}) and using the Bianchi identity 
$\nabla^\mu G_{\mu\nu} = 0$ and the conservation law $\nabla^\mu T_{\mu\nu} = 0$, we have the equation
\begin{eqnarray}
\nabla_\mu ( M_{Pl}^2 R + T ) = 0,
\label{UG2}
\end{eqnarray}
from which we can conclude that $M_{Pl}^2 R + T$ is a mere constant. If we define this integration constant as
\begin{eqnarray}
M_{Pl}^2 R + T = 4 \Lambda,
\label{UG3}
\end{eqnarray}
and insert this equation to Eq.  (\ref{UG}), we can obtain the standard Einstein equation with a cosmological 
constant $\Lambda$\footnote{Precisely speaking, the conventional Einstein equation with a cosmological
constant $\Lambda_0$ is described as $G_{\mu\nu} + \Lambda_0 g_{\mu\nu} = \frac{1}{M_{Pl}^2}T_{\mu\nu}$.
In this article, we have defined the cosmological constant as $\Lambda = M_{Pl}^2 \Lambda_0$, which is
nothing but the vacuum energy density usually written as $\rho$.}
\begin{eqnarray}
M_{Pl}^2 G_{\mu\nu} + \Lambda g_{\mu\nu} = T_{\mu\nu}.
\label{Standard E-eq}
\end{eqnarray}

In the above "derivation" of the equation of motion (\ref{Filter-Eins-Eq3}), we have considered the large length
limit $L \rightarrow \infty$, by which the filter excises only the infinite wave length and period fluctuations from
dynamics \cite{Nima}. This prescription of taking the limit $L \rightarrow \infty$ corresponds to taking the 
Fourier transform at vanishing frequency and momenta, which is in essence equivalent to taking 
the space-time average of dynamical observables. Comparing Eq. (\ref{Filter-Eins-Eq2}) with Eq. (\ref{Standard E-eq}), 
we find that the constant $c$ corresponds to the cosmological constant $\Lambda$. 
Since the cosmological constant is a global variable, which is a parameter of a system of codimension zero, 
and its measurement requires separating it from all the other long wave length modes in the universe, 
it takes a detector of the size of the universe to measure the value of the cosmological constant \cite{Kaloper1}. 
In this sense, taking the space-time average of the constant $c$ in Eq. (\ref{c-rel2}) makes sense as well.   

The equation of motion obtained in Eq. (\ref{Filter-Eins-Eq3}) has a remarkable property in that it $\it{sequesters}$
the vacuum energy from matter fields at both classical and quantum levels. To show this fact explicitly, let us
write the energy-momentum tensor 
\begin{eqnarray}
T_{\mu\nu} = - V_{vac} g_{\mu\nu} + \tau_{\mu\nu},
\label{Stress tensor}
\end{eqnarray}
where $V_{vac}$ denotes the vacuum energy density involving a classical cosmological constant and quantum 
fluctuations of the matter fields while $\tau_{\mu\nu}$ is the local excitation involving radiation and non-relativistic
matters etc. Because of the structure of the RHS of Eq. (\ref{Filter-Eins-Eq3}), the vacuum energy density
$V_{vac}$ completely decouples from the modified Einstein equation Eq. (\ref{Filter-Eins-Eq3}) since we have
\begin{eqnarray}
M_{Pl}^2 G_{\mu\nu} + \frac{1}{4} M_{Pl}^2 \overline{R} g_{\mu\nu} = \tau_{\mu\nu} - \frac{1}{4} \overline{\tau} g_{\mu\nu}.
\label{Filter-Eins-Eq4}
\end{eqnarray}
As a result, we have a residual effective cosmological constant given by
\begin{eqnarray}
\Lambda_{eff} \equiv \frac{1}{4} M_{Pl}^2 \overline{R} + \frac{1}{4} \overline{\tau}.
\label{Res-Cosmo}
\end{eqnarray}
Note that this quantity is stable under radiative corrections and would in general take a small value in a large and old universe.   
  
To close this section, we should comment on the operation of taking the space-time average in more detail. Obviously, the definition 
of this operation is very sutble in case of the open universes with infinite space-time volume, $V \equiv \int d^4 x \sqrt{-g} = \infty$. 
Since the residual cosmological constant 
includes $\overline{R}$ and $\overline{\tau}$ in the definition (\ref{Res-Cosmo}), we will focus on the two quantities 
in what follows.\footnote{The physical meaning of $\overline{R}$ has been already mentioned in Ref. \cite{Nima}.}   
First of all, let us notice that in any space-time with infinite volume, when $\int d^4 x \sqrt{-g} R$ is finite, $\overline{R}$
takes the vanishing value. This situation includes the physically interesting space-times where the universe is described 
in terms of a radiation or matter-dominated Friedmann-Robertson-Walker (FRW) universe. As a different example,  let us consider 
the universe which begins with the big bang and then is approaching the Sitter or Minkowski space-time asymptotically. 
Then, both the numerator and denominator in $\overline{R}$ are completely dominated  by the infinite 
contributions from the asymptotic space-time region in the future, so $\overline{R}$ is plausibly supposed to be equal to 
its asymptotic value $\overline{R} = R_{\infty}$ \cite{Nima}.   

Next, as for $\overline{\tau}$, in the proccess of expansion of the universe, it is natural to assume that the matter density
is gradually dilued in such a way that in the open universes $\overline{\tau}$ associated with the local excitations would be
asymptotically vanishing,  $\overline{\tau} =0$. Consequenly, in these space-times the residual cosmological constant would 
take the value 
\begin{eqnarray}
\Lambda_{eff} \equiv \frac{1}{4} M_{Pl}^2 R_{\infty}.
\label{Res-Cosmo2}
\end{eqnarray}
Thus, in space-times with $\overline{R} = 0$ or $R_{\infty} \sim 0$ like an asymptotically Minkowskian space-time, the late-time acceleration 
cannot be explained in the theory at hand. Moreover, it is highly improbable that Eq. (\ref{Res-Cosmo2}) happens to provide precisely 
the current value $\sim (1 meV)^4$ under a situation without physically plausible reasons. 
It therefore seems that we need to find a more sophisticated mechanism into consideration in order to account for the late-time 
acceleration of the universe when the the formula (\ref{Res-Cosmo2}) does not match the current value of the cosmological constant
or takes the vanishing result.  
We wish to present such a new mechanism in Section 4 on the basis of the formulation explained in the next section.
Before doing it, let us briefly review our formulation of the nonlocal approach to the cosmological constant \cite{Oda7}.

\section{Review of the $R^2$ model}

In this section, we will review a manifestly local and generally coordinate invariant formulation \cite{Carroll}-\cite{Oda2} 
for a nonlocal approach to the cosmological constant problem, in particular, our most recent work \cite{Oda7}.\footnote{See 
the related papers \cite{Oda4}-\cite{Oda6}.}  

A manifestly local and generally coordinate invariant action for our nonlocal approach consists of two parts
\begin{eqnarray}
S = S_{GR} + S_{Top},
\label{T-Action}
\end{eqnarray}
where the gravitational action $S_{GR}$ with a bare cosmological constant $\Lambda$ as well as a generic matter Lagrangian density 
${\cal{L}}_{m} = {\cal{L}}_{m} (g_{\mu\nu}, \Psi)$ ($\Psi$ is a generic matter field), and the topological action $S_{Top}$ are 
respectively defined as
\begin{eqnarray}
S_{GR}  = \int d^4 x \sqrt{- g} \ \Biggl[ \eta(x) ( R - 2 \Lambda ) + {\cal{L}}_{m} 
- \frac{1}{2} \cdot \frac{1}{4!} F_{\mu\nu\rho\sigma}^2 
+ \frac{1}{6} \nabla_\mu ( F^{\mu\nu\rho\sigma} A_{\nu\rho\sigma} )
\Biggr],
\label{GR-Action}
\end{eqnarray}
and
\begin{eqnarray}
S_{Top}  = \int d^4 x  \ \frac{1}{4!}  \ \mathring{\varepsilon}^{\mu\nu\rho\sigma} M_{Pl}^2
f\left( \frac{\eta(x)}{M_{Pl}^2}  \right) H_{\mu\nu\rho\sigma},
\label{Top-Action}
\end{eqnarray}
where $\eta(x)$ is a scalar field of dimension of mass squared and $f(x)$ is a function which cannot be a linear function. 
Let us note that the scalar field  $\eta(x)$ has no local degrees of freedom except the zero mode because of 
the gauge symmetry of the 4-form strength \cite{Henneaux}.  
Moreover, $F_{\mu\nu\rho\sigma}$ and $ H_{\mu\nu\rho\sigma}$ are respectively the field strengths 
for two 3-form gauge fields $A_{\mu\nu\rho}$ and $B_{\mu\nu\rho}$
\begin{eqnarray}
F_{\mu\nu\rho\sigma} = 4 \partial_{[\mu} A_{\nu\rho\sigma]},  \qquad 
H_{\mu\nu\rho\sigma} = 4 \partial_{[\mu} B_{\nu\rho\sigma]},
\label{4-forms}
\end{eqnarray}
where the square brackets denote antisymmetrization of enclosed indices. Finally, $\mathring{\varepsilon}^{\mu\nu\rho\sigma}$
and $\mathring{\varepsilon}_{\mu\nu\rho\sigma}$ are the Levi-Civita tensor density defined as
\begin{eqnarray}
\mathring{\varepsilon}^{0123} = + 1,  \qquad 
\mathring{\varepsilon}_{0123} = - 1,
\label{Levi}
\end{eqnarray}
and they are related to the totally antisymmetric tensors $\varepsilon^{\mu\nu\rho\sigma}$ and $\varepsilon_{\mu\nu\rho\sigma}$ via
relations
\begin{eqnarray}
\varepsilon^{\mu\nu\rho\sigma} = \frac{1}{\sqrt{-g}}  \mathring{\varepsilon}^{\mu\nu\rho\sigma},  \qquad 
\varepsilon_{\mu\nu\rho\sigma} = \sqrt{-g}  \mathring{\varepsilon}_{\mu\nu\rho\sigma}.
\label{Levi-tensor}
\end{eqnarray}

Now let us derive all the equations of motion from the action (\ref{T-Action}).  First of all, the variation 
with respect to the 3-form $B_{\mu\nu\rho}$ gives rise to the equation for a scalar field  $\eta(x)$:
\begin{eqnarray}
\mathring{\varepsilon}^{\mu\nu\rho\sigma} f^\prime \partial_\sigma \eta(x)  = 0,
\label{B-eq}
\end{eqnarray}
where $f^\prime (x) \equiv \frac{d f(x)}{dx}$.  From this equation, we have a classical solution for $\eta(x)$:
\begin{eqnarray}
\eta(x)  = \eta,
\label{eta-sol}
\end{eqnarray}
where $\eta$ is a certain constant.  Next, taking the variation of the scalar field $\eta(x)$ leads to the
equation:
\begin{eqnarray}
\sqrt{- g} ( R - 2 \Lambda ) + \frac{1}{4!}  \ \mathring{\varepsilon}^{\mu\nu\rho\sigma} f^\prime H_{\mu\nu\rho\sigma} = 0.
\label{eta-equation}
\end{eqnarray}
Since we can always set $H_{\mu\nu\rho\sigma}$ to be
\begin{eqnarray}
H_{\mu\nu\rho\sigma}  = c(x) \varepsilon_{\mu\nu\rho\sigma} = c(x) \sqrt{-g} \mathring{\varepsilon}_{\mu\nu\rho\sigma},
\label{H-equation}
\end{eqnarray}
with $c(x)$ being some scalar function, Eq.  (\ref{eta-equation}) can be rewritten as
\begin{eqnarray}
R - 2 \Lambda - f^\prime c(x) = 0.
\label{H-equation 2}
\end{eqnarray}
In order to take account of the cosmological constant problem, let us take the space-time average of this equation,
which provides a constraint equation:
\begin{eqnarray}
\overline{R} = 2 \Lambda + f^\prime \left( \frac{\eta}{M_{Pl}^2} \right) \overline{c(x)},
\label{H-constraint}
\end{eqnarray}
where we have used Eq. (\ref{eta-sol}).

The equation of motion for the 3-form $A_{\mu\nu\rho}$ gives the Maxwell-like equation:
\begin{eqnarray}
\nabla^\mu F_{\mu\nu\rho\sigma} = 0.
\label{A-eq}
\end{eqnarray}
As in $H_{\mu\nu\rho\sigma}$, if we set 
\begin{eqnarray}
F_{\mu\nu\rho\sigma}  = \theta(x) \varepsilon_{\mu\nu\rho\sigma} = \theta(x) \sqrt{-g} 
\mathring{\varepsilon}_{\mu\nu\rho\sigma},
\label{F}
\end{eqnarray}
with $\theta(x)$ being a scalar function, Eq. (\ref{A-eq}) requires $\theta(x)$ to be a mere constant
\begin{eqnarray}
\theta(x) = \theta,
\label{Theta}
\end{eqnarray}
where $\theta$ is a constant. 

Finally, the variation with respect to the metric tensor yields the gravitational field equation, i.e., the Einstein equation:
\begin{eqnarray}
\eta \left(  G_{\mu\nu}  + \Lambda g_{\mu\nu} \right)  - \frac{1}{2} T_{\mu\nu}
+ \frac{1}{4} \cdot \frac{1}{4!} g_{\mu\nu} F_{\alpha\beta\gamma\delta}^2 
- \frac{1}{12} F_{\mu\alpha\beta\gamma} F_\nu \,^{\alpha\beta\gamma}
= 0,
\label{Eins-eq 1}
\end{eqnarray}
where the energy-momentum tensor is defined by $ T_{\mu\nu} = - \frac{2}{\sqrt{-g}} \frac{\delta (\sqrt{-g}  {\cal{L}}_{m})}
{\delta g^{\mu\nu}}$.  In deriving Eq. (\ref{Eins-eq 1}), we have again used Eq. (\ref{eta-sol}).
Furthermore, using Eqs. (\ref{F}) and (\ref{Theta}), this equation can be simplified to be the form
\begin{eqnarray}
\eta \left(  G_{\mu\nu}  + \Lambda g_{\mu\nu} \right)  - \frac{1}{2} T_{\mu\nu}
+ \frac{1}{4} \theta^2 g_{\mu\nu} = 0.
\label{Eins-eq 1-theta}
\end{eqnarray}
Taking the trace of Eq.  (\ref{Eins-eq 1-theta}) and then the space-time average, one obtains a constraint
\begin{eqnarray}
\theta^2 = \frac{1}{2} \overline{T} + \eta ( \overline{R} - 4 \Lambda).
\label{Eins-constraint}
\end{eqnarray}
Substituting this constraint into the Einstein equation (\ref{Eins-eq 1-theta}), we find that
\begin{eqnarray}
M_{Pl}^2 G_{\mu\nu}  + \frac{1}{4} M_{Pl}^2 \overline{R}  g_{\mu\nu} = T_{\mu\nu}
- \frac{1}{4} \overline{T} g_{\mu\nu},
\label{Final Eins-eq 1}
\end{eqnarray}
where we have chosen $\eta = \frac{M_{Pl}^2}{2}$. Note that this gravitational equation precisely coincides 
with the equation of motion (\ref{Filter-Eins-Eq3}), which was obtained from the filter mechanism.
In other words, the action (\ref{T-Action}) provides us with a concrete realization of the gravitational model
with a high-pass filter obtained in a rather phenomenological manner in Section 2.  

As done in Section 2, if we separate the energy-momentum tensor $T_{\mu\nu}$ into two parts as in 
Eq. (\ref{Stress tensor}), we have the equation of motion for the gravitational field, Eq. (\ref{Filter-Eins-Eq4}),
with the residual effective cosmological constant  (\ref{Res-Cosmo}).
The first term in the RHS of Eq. (\ref{Res-Cosmo}) is radiatively stable since $\overline{R}$ is so.
Actually, as seen in Eq.  (\ref{H-constraint}), $\Lambda$ is a mere number and $\overline{c(x)}$
is proportional to the flux of the 4-form which is the infrared (IR) quantity, thereby implying that 
$\overline{R}$ is a radiatively stable quantity. The second term $\frac{1}{4} 
\overline{\tau}$ in the RHS of Eq. (\ref{Res-Cosmo}) is obviously radiatively stable. Thus, our cosmological 
constant $\Lambda_{eff}$ is a radiatively stable quantity so it can be fixed by the measurement in a consistent manner.
 
The only disadvantage of this nonlocal approach to the cosmological constant problem is that
we confine ourselves to the semiclassical approach where the matter loop effects are included
in the energy-momentum tensor while the quantum gravity effects are completely ignored.
In this context, note that gravity is a classical field merely serving the purpose of detecting the vacuum energy.

Next,  let us therefore turn our attention to quantum gravity effects \cite{Oda7}.
From the 1-loop calculation, the dimensional analysis and general covariance, it is easy to estimate the loop 
effects from both matter and the gravitational fields.  For instance, the renormalization of the Newton constant 
and the cosmological constant amounts to adding the following effective action to the total action (\ref{T-Action})
up to the logarithmic divergences which are so subdominant that they are irrelevant to the argument at hand \cite{Kaloper3}:
\begin{eqnarray}
S_q  &=& \int d^4 x \sqrt{- g} \ \Biggl[ \left( a_0 M^2 + a_1 \frac{M^4}{\eta} + a_2 \frac{M^6}{\eta^2}
+ \cdots \right) R + b_0 M^4 + b_1 \frac{M^6}{\eta} + b_2 \frac{M^8}{\eta^2} + \cdots \Biggr]
\nonumber\\
&\equiv& \int d^4 x \sqrt{- g} \ \left[ \alpha (\eta) R + \beta(\eta) \right],
\label{q-Action}
\end{eqnarray}
where $M$ is a cutoff and the coefficients $a_i, b_i ( i = 0, 1, 2, \cdots )$ are ${\cal{O}}(1)$.  
As shown in Ref. \cite{Oda7}, when we start with the action $S = S_{GR} + S_{Top} + S_q$, we find that
the effective cosmological constant $\Lambda_{eff}$ in Eq. (\ref{Res-Cosmo}) is not radiatively stable because 
of the presence of $\beta(\eta)$, which means that the graviton loop effects render the effective cosmological constant
be radiatively unstable.

To remedy this situation, we have added the higher-derivative $R^2$ term and the corresponding topological
term \cite{Oda7}. Then, the total effective action constituting of four parts is given by
\begin{eqnarray}
S = S_{GR} + S_{Top} + S_q + S_{R^2},
\label{T-R2-Action}
\end{eqnarray}
where the last action $S_{R^2}$ is defined as
\begin{eqnarray}
S_{R^2}  = \int d^4 x \sqrt{- g} \omega(x) R^2 + \int d^4 x  \ \frac{1}{4!}  \ \mathring{\varepsilon}^{\mu\nu\rho\sigma} 
M_{Pl}^2 \hat f(\omega) \hat H_{\mu\nu\rho\sigma},
\label{R2-Action}
\end{eqnarray}
where $\hat H_{\mu\nu\rho\sigma} \equiv 4 \partial_{[\mu} \hat B_{\nu\rho\sigma]}$. 

As before, the variation of the total action (\ref{T-R2-Action}) with respect to the 3-form $\hat B_{\mu\nu\rho}$ produces
\begin{eqnarray}
\mathring{\varepsilon}^{\mu\nu\rho\sigma} \hat f^\prime \partial_\sigma \omega(x)  = 0,
\label{hat-B-eq}
\end{eqnarray}
which gives us a classical solution for $\omega(x)$:
\begin{eqnarray}
\omega(x) = \omega,
\label{omega-sol}
\end{eqnarray}
where $\omega$ is a certain constant.  Next, taking the variation of the scalar field $\omega(x)$ yields the
field equation:
\begin{eqnarray}
\sqrt{- g} R^2 + \frac{1}{4!}  \ \mathring{\varepsilon}^{\mu\nu\rho\sigma} M_{Pl}^2 \hat f^\prime 
\hat H_{\mu\nu\rho\sigma} = 0.
\label{omega-equation}
\end{eqnarray}
Setting $\hat H_{\mu\nu\rho\sigma}$ again to be
\begin{eqnarray}
\hat H_{\mu\nu\rho\sigma}  = \hat c(x) \varepsilon_{\mu\nu\rho\sigma} 
= \hat c(x) \sqrt{-g} \mathring{\varepsilon}_{\mu\nu\rho\sigma},
\label{hat-H-equation}
\end{eqnarray}
with $\hat c(x)$ being some scalar function, Eq.  (\ref{omega-equation}) can be cast to
\begin{eqnarray}
R^2 - M_{Pl}^2 \hat f^\prime \hat c(x) = 0.
\label{hat-H-equation 2}
\end{eqnarray}
Then, the space-time average of this equation leads to a new constraint equation:
\begin{eqnarray}
\overline{R^2} = M_{Pl}^2 \hat f^\prime (\omega) \overline{\hat c(x)}.
\label{hat-H-constraint}
\end{eqnarray}
Note that $\overline{R^2}$ is radiatively stable since both $\hat f^\prime (\omega)$ and $ \overline{\hat c(x)}$
are radiatively stable. 

With the help of Eqs. (\ref{eta-sol}),  (\ref{F}), (\ref{Theta}) and (\ref{omega-sol}), the variation with respect to 
the metric tensor yields the Einstein equation:
\begin{eqnarray}
\eta \left(  G_{\mu\nu}  + \Lambda g_{\mu\nu} \right)  + \omega \ {}^{(1)} H_{\mu\nu} 
+ \alpha(\eta) G_{\mu\nu} - \frac{1}{2} \beta(\eta) g_{\mu\nu} - \frac{1}{2} T_{\mu\nu}
+ \frac{1}{4} \theta^2 g_{\mu\nu} = 0,
\label{Grav-theta}
\end{eqnarray}
where ${}^{(1)} H_{\mu\nu}$ is defined as
\begin{eqnarray}
{}^{(1)} H_{\mu\nu} &=&  \frac{1}{\sqrt{-g}} \frac{\delta}{\delta g^{\mu\nu}} \int d^n x \sqrt{-g} R^2
\nonumber\\
&=& -2 \nabla_\mu \nabla_\nu R + 2 g_{\mu\nu} \Box R - \frac{1}{2} g_{\mu\nu} R^2 + 2 R R_{\mu\nu}.
\label{Def H1}
\end{eqnarray}
Following the same line of the argument as before, it is easy to arrive at the final form of the Einstein equation:
\begin{eqnarray}
M_{eff}^2 G_{\mu\nu}  + 2 \omega \ {}^{(1)} H_{\mu\nu} + \frac{1}{4} M_{eff}^2 \overline{R}  g_{\mu\nu} 
= \tau_{\mu\nu} - \frac{1}{4} \overline{\tau} g_{\mu\nu},
\label{Eins-eq 3}
\end{eqnarray}
where $M_{eff}^2 \equiv 2 ( \eta + \alpha(\eta)) = M_{Pl}^2 + 2 \alpha(\eta)$. This equation shows that 
the effective cosmological constant is again of the form
\begin{eqnarray}
\Lambda_{eff} =  \frac{1}{4} M_{eff}^2 \overline{R}  + \frac{1}{4} \overline{\tau}.
\label{CC-3}
\end{eqnarray}

Since the second term in the RHS is obviously radiatively stable, let us focus on the first term:
\begin{eqnarray}
\Delta \Lambda \equiv  \frac{1}{4} M_{eff}^2 \overline{R}.
\label{CC-4}
\end{eqnarray}
Then, the Einstein equation (\ref{Eins-eq 3}) reads
\begin{eqnarray}
M_{eff}^2 G_{\mu\nu}  + 2 \omega \ {}^{(1)} H_{\mu\nu} +  \Delta \Lambda g_{\mu\nu} 
= \tau_{\mu\nu} - \frac{1}{4} \overline{\tau} g_{\mu\nu}.
\label{Eins-eq 3-2}
\end{eqnarray}
Taking the trace of this equation, one obtains
\begin{eqnarray}
M_{eff}^2 \left( 1 - \frac{12 \omega}{M_{eff}^2} \Box \right) R  
= 4 \left[ \Delta \Lambda - \frac{1}{4} ( \tau - \overline{\tau} ) \right],
\label{Tr-Eins-eq 3-2}
\end{eqnarray}
from which one can express the scalar curvature as
\begin{eqnarray}
R = \frac{4}{M_{eff}^2} \left[ \Delta \Lambda - \frac{\tau - \overline{\tau}}{4} 
+ \frac{1}{4} \frac{\Box}{\Box - \frac{M_{eff}^2}{12 \omega}}  \tau \right].
\label{Tr-Eins-eq 3-3}
\end{eqnarray}
Note that this expression reduces to Eq. (\ref{CC-4}) up to a surface term when one takes the space-time average,
which guarantees the correctness of our derivation. 

Taking the square of Eq. (\ref{Tr-Eins-eq 3-3}) and then the space-time average makes it possible to
describe $(\Delta \Lambda)^2$ in terms of $\overline{R^2}$ and $\tau$
\begin{eqnarray}
(\Delta \Lambda)^2 =  \frac{M_{eff}^4}{16} \overline{R^2} - \frac{1}{16} \left( \overline{\tau^2} 
- \overline{\tau}^2 \right) - \frac{1}{16} \overline{\left( \frac{\Box}{\Box - \frac{M_{eff}^2}{12 \omega}}  \tau
\right)^2 } 
+ \frac{1}{8} \overline{\left(\tau \frac{\Box}{\Box - \frac{M_{eff}^2}{12 \omega}} \tau \right)}.
\label{Delta Lambda}
\end{eqnarray}
This expression clearly shows that $\Delta \Lambda$ is radiatively stable since $M_{eff}^2$
turns out to be radiatively stable, $ \overline{R^2}$ is also radiatively stable as shown in Eq. (\ref{hat-H-constraint}),
and the remaining terms involving $\tau$ are radiatively stable as well as long as $\Box$ is not equal to 
$\frac{M_{eff}^2}{12 \omega}$.  The radiative stability of $\Delta \Lambda$ ensures that the effective
cosmological constant $\Lambda_{eff}$ in Eq. (\ref{CC-3}) is also radiatively stable even when the gravitational
radiative corrections are taken into account.

Before closing this section, we would like to comment on two remarks. One of them is that in the high energy limit, 
$(\Delta \Lambda)^2$ reduces to
\begin{eqnarray}
(\Delta \Lambda)^2 \rightarrow  \frac{M_{eff}^4}{16} \overline{R^2} + \frac{1}{16} \overline{\tau}^2,
\label{Delta Lambda 2}
\end{eqnarray}
which is manifestly radiatively stable. The other remark is that we see that using Eqs. (\ref{CC-4}) and 
(\ref{Delta Lambda}),  $\overline{R}$ is also radiatively stable whose situation should be contrasted with
the case without the $R^2$ term, for which $\overline{R}$ is not so.

\section{Cosmic acceleration} 
 
One might be concerned that the higher-derivative term $\omega R^2$, which was introduced in the action (\ref{R2-Action}),
would generate new radiative corrections that also depend on $\omega$, thereby inducing new radiative corrections to
the vacuum energy density and consequently breaking its radiative stability. We will first show that this is not indeed 
the case explicitly when the mass of the scalaron is around $1 meV$. Nevertheless, it turns out that the existence of
the scalaron enables the cosmological constant to have the currently observed value in the open universes.

For this purpose, it is convenient to move from the Jordan frame to the Einstein one since the higher-derivative 
$R^2$ gravity is conformally equivalent to Einstein gravity with an extra scalar field called "scalaron" in the Einstein frame. 
Of course, one should not confuse the conformal metric with the original physical metric since they generally describe manifolds
with different geometries and the final results should be interpreted in terms of the original metric. However, as seen shortly, 
it turns out that our conformal factor connecting with the two metrics depends on the scalar curvature which does not change 
significantly during the late-time acceleration, so we have same results in the both frames.  

With the help of Eqs. (\ref{eta-sol}), (\ref{F}), (\ref{Theta}) and (\ref{omega-sol}),  the total action 
(\ref{T-R2-Action}) reads    
\begin{eqnarray}
S  &=& \int d^4 x \sqrt{- g} \ \Biggl[ \eta ( R - 2 \Lambda ) + {\cal{L}}_{m} - \frac{1}{2} \theta^2 
+ \alpha(\eta) R + \beta(\eta) + \omega R^2  \Biggr]
\nonumber\\
&=& \int d^4 x \sqrt{- g} \ \Biggl[ \frac{M_{eff}^2}{2} R - \hat \Lambda + \omega R^2  + {\cal{L}}_{m} \Biggr],
\label{Cos-Action1}
\end{eqnarray}
where we have defined 
\begin{eqnarray}
\hat \Lambda =  2 \eta \Lambda + \frac{1}{2} \theta^2 - \beta(\eta).
\label{Hat-Lambda}
\end{eqnarray}
Now, using the conformal transformation \cite{Tsujikawa}
\begin{eqnarray}
\bar g_{\mu\nu} =  \left( 1 + \frac{4 \omega}{M_{eff}^2} R \right) g_{\mu\nu},
\label{Conf-transf}
\end{eqnarray}
we find that the action (\ref{Cos-Action1}) takes the form in the Einstein frame
\begin{eqnarray}
S  = \int d^4 x \sqrt{- \bar g} \ \Biggl[ \frac{M_{eff}^2}{2} \bar R - \frac{1}{2} \bar g^{\mu\nu} 
\partial_\mu \phi \partial_\nu \phi - V (\phi) + e^{- 2 \sqrt{\frac{2}{3}} \frac{\phi}{M_{eff}}}
{\cal{L}}_{m} (e^{- \sqrt{\frac{2}{3}} \frac{\phi}{M_{eff}}} \bar g_{\mu\nu}, \Psi) \Biggr],
\label{Cos-Action2}
\end{eqnarray}
where we have defined a scalar field $\phi$ and the potential $V (\phi)$ as
\begin{eqnarray}
\phi  &=& \sqrt{\frac{3}{2}} M_{eff} \log \left( 1 +  \frac{4 \omega}{M_{eff}^2} R \right),
\nonumber\\
V (\phi) &=& \frac{M_{eff}^4}{16 \omega} \left( 1 - e^{-  \sqrt{\frac{2}{3}} \frac{\phi}{M_{eff}}} \right)^2
+  \hat \Lambda e^{- 2 \sqrt{\frac{2}{3}} \frac{\phi}{M_{eff}}}.
\label{phi-V}
\end{eqnarray}

Since we take account of the late-time acceleration, it is sufficient to confine ourselves to the low curvature regime
\begin{eqnarray}
\phi \ll M_{eff}.
\label{Low-R}
\end{eqnarray}
In this regime, the scalaron $\phi$ decouples in the matter Lagrangian:
\begin{eqnarray}
e^{- 2 \sqrt{\frac{2}{3}} \frac{\phi}{M_{eff}}}
{\cal{L}}_{m} (e^{- \sqrt{\frac{2}{3}} \frac{\phi}{M_{eff}}} \bar g_{\mu\nu}, \Psi)
\sim {\cal{L}}_{m} (\bar g_{\mu\nu}, \Psi).
\label{Matt-Lag}
\end{eqnarray}
Furthermore, the potential $V (\phi)$ can be expanded in the Taylor series as
\begin{eqnarray}
V (\phi) = \hat \Lambda - 2 \sqrt{\frac{2}{3}} \frac{\hat \Lambda}{M_{eff}} \phi
+ \left( \frac{4}{3} \frac{\hat \Lambda}{M_{eff}^2} + \frac{M_{eff}^2}{24 \omega} \right) \phi^2
- \sqrt{\frac{2}{3}} \left( \frac{8}{9} \frac{\hat \Lambda}{M_{eff}^3} 
+ \frac{M_{eff}}{24 \omega} \right) \phi^3 + {\cal{O}} (\phi^4).
\label{V-exp}
\end{eqnarray}

It has been recently shown in Ref. \cite{Brax} that the compatibility between the acceleration of the expansion rate 
of the universe, local tests of gravity and the quantum stability of the theory converges to select the relation
\begin{eqnarray}
\omega \sim \frac{M_{Pl}^2}{\sqrt{\hat \Lambda}} \sim \frac{M_{eff}^2}{\sqrt{\hat \Lambda}}
\sim 10^{61}.
\label{Relation}
\end{eqnarray}
With this relation, the potential $V(\phi)$ can be approximated by
\begin{eqnarray}
V (\phi) \sim \hat \Lambda - 2 \sqrt{\frac{2}{3}} \frac{\hat \Lambda}{M_{eff}} \phi
+ \frac{M_{eff}^2}{24 \omega} \phi^2
- \sqrt{\frac{2}{3}} \frac{M_{eff}}{24 \omega} \phi^3 + {\cal{O}} (\phi^4).
\label{V-exp2}
\end{eqnarray}
Using this approximated potential, we can obtain the minimum $\phi_\star$ for the scalaron and its effective mass 
$m_\star$\footnote{Here we neglect the chameleon mechanism \cite{Khoury}, which is irrelevant to the present
discussion.}
\begin{eqnarray}
\phi_\star \sim 24 \sqrt{\frac{2}{3}} \frac{\hat \Lambda}{M_{eff}^3} \omega,   \qquad
m_\star^2 \sim \frac{M_{eff}^2}{12 \omega}.
\label{Star}
\end{eqnarray}
Note that using Eqs. (\ref{Relation}) and (\ref{Star}), we find that the effective scalaron mass is given by
\begin{eqnarray}
m_\star \sim 1 meV.
\label{m-star}
\end{eqnarray}

Then, using the formula (\ref{Renorm-Lambda-1 loop}), the 1-loop renormalized cosmological constant
up to a finite part is easily calculated to be
\begin{eqnarray}
\Lambda^{\phi, 1-loop}_{ren} = \frac{m_\star^4}{(8 \pi)^2} \log \left( \frac{m_\star^2}{M^2} \right)
\sim \frac{M_{eff}^4}{(96 \pi)^2 \omega^2} \log \left( \frac{M_{eff}^2}{12 \omega M^2} \right),
\label{Renorm-Lambda-1 loop2}
\end{eqnarray}
which implies that\footnote{We have selected $M$ to be the current Hubble constant, i.e., 
$M \sim 10^{-30} meV$.}
\begin{eqnarray}
\Lambda^{\phi, 1-loop}_{ren} \sim (1 meV)^4.
\label{Renorm-Lambda-1 loop3}
\end{eqnarray}
We have therefore obtained the current value of the cosmological constant. The important point is that 
owing to the form of the potential (\ref{V-exp2}), the higher-order contributions to the cosmological constant are 
strictly suppressed by more $\frac{1}{\omega}$ factors, so these contributions are very tiny compared to the 1-loop result,
thereby implying that the cosmological constant in the theory at hand is radiatively stable even if the $R^2$ term and its radiative 
corrections are incorporated.

\section{Discussions}

In our previous work \cite{Oda7}, we have constructed a nonlocal approach to the cosmological constant 
problem which includes both gravity and matter loop effects. In this approach, we are naturally led to
incorporate the higher-derivative $R^2$ term in the action in order to guarantee the radiative stability of the cosmological 
constant under the gravitational loop effects.  The presence of such a higher-derivative term generally generates 
additional radiative corrections which depend on the coefficient, that is, the coupling constant in front of 
the higher-derivative term.  It was suggested in the previous paper \cite{Oda7} that such a renormalizable term 
would keep the radiative stability when the mass of the scalaron is very tiny.  In the present paper, we have explicitly shown 
this fact by moving from the Jordan frame to the Einstein one, calculate the 1-loop renormalized cosmological constant, 
and then examine higher-loop corrections via the analysis of interaction terms in the low curvature regime.

In cosmology, it is known that the $R^2$ gravity \cite{Staro} and its generalization, $f(R)$ gravity
\cite{Tsujikawa}, play a role in various cosmological phenomena such as inflation, dark energy and cosmological 
perturbations etc. In this article, we have therefore investigated a possibility of applying our nonlocal approach to 
the cosmological constant problem having the $R^2$ term \cite{Oda7} for the late-time acceleration of the expansion 
of the universe. So far, the similar formulations have been already applied to inflation in the early universe \cite{Kaloper1, Avelino}, 
but these formulations ignore quantum effects coming from the graviton loops so that they are not free from the radiative 
instability issue. On the other hand, since our formulation includes both gravity and matter loop effects, the cosmological constant 
is completely stable under radiative corrections. 

In the nonlocal approach to the cosmological constant problem \cite{Carroll, Oda1, Oda7} or the model of vacuum energy sequestering 
\cite{Kaloper1, Kaloper2, Kaloper3, Kaloper4}, the residual effective cosmological constant is automatically very small in a large and old
universe. The main advantage of this formalism is that all the vacuum energy corrections except a finite and tiny residual cosmological
constant are completely removed from the gravitational equation. In particular, in the open universes with infinite space-time
volume, it seems to be natural to conjecture that the residual cosmological constant is exactly zero or very tiny compared to the
present value of the cosmological constant. In other words, the present mechanism seems to sequester away the vacuum energy
almost completely in case of the open universes.

Then, it is a natural question to ask ourselves if the nonlocal approach to the cosmological constant problem \cite{Oda7} is compatible
with inflation at an early epoch or the late-time acceleration of the universe. In this article, we have answered this question partially 
that our formalism can be applied to the late-time acceleration where the key observation is that the present mechanism cannot sequester 
away the vacuum energy associated with the scalaron existing in the $R^2$ gravity as an extra scalar field. When the mass of the
scalaron is about $1 meV$, which is strongly supported by the recent analysis in \cite{Brax}, our formalism not only accounts for 
the current cosmological constant given by $(1 meV)^4$ but also assures the radiative stability of the vacuum energy density.

It is worthwhile to mention the relation between the present work and the recent work \cite{Brax}.  In Ref. \cite{Brax}, 
it is explicitly assumed that all matter contributions to the vacuum energy vanish or contribute in a negligible way to the vacuum energy at
the energy scale $\mu$ associated with our universe.\footnote{For simplicity, we neglect the classical energy density in the gravity
sector.} Thus, the following tuning to the vacuum energy is assumed:
\begin{eqnarray}
2 \Lambda_0 M_{Pl}^2 + \rho_{trans} + \sum_j (-1)^{2j} ( 2j+1) \theta(m_j - \mu_0) \frac{m_j^4}{(8 \pi)^2} 
\log \frac{m_j^2}{\mu^2} = 0,
\label{Brax}
\end{eqnarray}
where the first term is the bare cosmological constant, $\rho_{trans}$ denotes the energy density corresponding to
all the phase transitions such as the electro-weak phase transition, $m_j$ is the mass of a particle of spin $j$, and
the mass scale $\mu_0$ is for instance chosen to be the electron mass, which is the minimum mass of observed particles
thus far, apart from the neutrinos which might be
of the right order of magnitude if the neutrino masses are in the order $1 meV$. The appealing point of our nonlocal
approach to the cosmological constant \cite{Oda7} is that this equation (\ref{Brax}) is derived in a natural manner.
Furthermore, radiative corrections from gravitational field are also sequestered away.

Finally, let us comment on a future work. The important problem is to explain the inflation at the early epoch within 
the present framework. In the inflationary scenario, it is postulated that the universe is dominated by a transient large vacuum 
energy. However, our nonlocal approach usually removes the large vacuum energy and leaves only small one.
At present, there seems to be a difficulty to reconcile the both results without losing the radiative stability.
In any case, we wish to return to this problem in future.

\begin{flushleft}
{\bf Acknowledgements}
\end{flushleft}

This work was supported by JSPS KAKENHI Grant Number 16K05327.



\begin{thebibliography}{99}

\bibitem{Riess}
A. G. Riess et al. [Supernova Search Team], {Astron. J. {\bf 116} (1998) 1009.}

\bibitem{Perlmutter}
S. Perlmutter et al. [Supernova Cosmology Project Collaboration], {Astrophys. J. {\bf 517} (1999) 565.}

\bibitem{Hinshaw}
G. Hinshaw et al. [WMAP Collaboration], {Astrophys. J. Suppl. Ser. {\bf 208} (2013) 19.}

\bibitem{Ade}
P. A. R. Ade et al. [Planck Collaboration], {Astron. Astrophys. {\bf A 13} (2016) 594.}

\bibitem{Guth}
A. H. Guth, {Phys. Rev. {\bf D 23} (1981) 347.}

\bibitem{Sato}
K. Sato, {Mon. Not. Roy. Astron. Soc. {\bf 195} (1981) 467.}

\bibitem{Linde1}
A. D. Linde, {Phys. Lett. {\bf B 108} (1982) 389.}

\bibitem{Albrecht}
A. Albrecht and P. J. Steinhardt, {Phys. Rev. Lett. {\bf 48} (1982) 1220.}

\bibitem{Staro}
A. A. Starobinsky, {Phys. Lett. {\bf B 91} (1980) 99.}

\bibitem{Weinberg}
S. Weinberg, {Rev. Mod. Phys. {\bf 61} (1989) 1.}

\bibitem{Padilla}
A. Padilla, {arXiv:1502.05296 [hep-th].}

\bibitem{Oda7}
I. Oda, {arXiv:1709.08189 [hep-th], Phys. Rev. {\bf D} (in press).}  

\bibitem{MTW}
C. W. Misner, K. S. Thorne and J. A. Wheeler, {"Gravitation", W H Freeman and Co (Sd), 1973.}

\bibitem{Brax}
P. Brax, P. Valageas and P. Vanhove, {arXiv:1711.03356 [astro-ph.CO].}

\bibitem{Linde}
A. D. Linde, {Phys. Lett. {\bf B 200} (1988) 272.}

\bibitem{Tseytlin}
A. A. Tseytlin, {Phys. Rev. Lett. {\bf 66} (1991) 545.}

\bibitem{Nima}
N. Arkani-Hamed, S. Dimopoulos, G. Dvali and G. Gabadadze, {arXiv: hep-th/0209227.}

\bibitem{Kaloper1}
N. Kaloper and A. Padilla, {Phys. Rev. Lett. {\bf 112} (2014) 091304;
Phys. Rev. {\bf D 90} (2014) 084023; Phys. Rev. Lett. {\bf 114} (2015) 101302.}

\bibitem{Kaloper2}
N. Kaloper, A. Padilla, D. Stefanyszyn and G. Zahariade, {Phys. Rev. Lett. {\bf 116} (2016) 051302.}

\bibitem{Kaloper3}
N. Kaloper and A. Padilla, {Phys. Rev. Lett. {\bf 118} (2017) 061303.}

\bibitem{Carroll}
S. M. Carroll and G. N. Remmen, {Phys. Rev. {\bf D 95} (2017) 123504.} 

\bibitem{Oda1}
I. Oda, {Phys. Rev. {\bf D 95} (2017) 104020.}

\bibitem{Kaloper4}
G. D'Amico, N. Kaloper, A. Padilla, D. Stefanyszyn, A. Westphal and G. Zahariade, {JHEP {\bf 09} (2017) 074.}

\bibitem{Oda2}
I. Oda, {Phys. Rev. {\bf D 96} (2017) 024027.}

\bibitem{Oda3}
I. Oda, {arXiv:1706.05804 [hep-th].}  

\bibitem{Avelino}
P. P. Avelino, {Phys. Rev. {\bf D 90} (2014) 103523.}

\bibitem{Linde2}
A. D. Linde, {Mod. Phys.  Lett. {\bf A 01} (1986) 81.}

\bibitem{Einstein}
A. Einstein, {in {"The Principle of Relativity", by A. Einstein et al., Dover
Publications, New York, 1952.}}

\bibitem{Oda4}
I. Oda, {PTEP {\bf 2016} (2016) 081B01.}

\bibitem{Oda5}
I. Oda, {Adv. Studies in Theor. Phys. {\bf 10} (2016) 319.} 

\bibitem{Oda6}
I. Oda, {Int. J. Mod. Phys. {\bf D 26} (2016) 1750023.}

\bibitem{Henneaux}
M. Henneaux and C. Teitelboim, {Phys. Lett. {\bf B 222} (1989) 195.}

\bibitem{Tsujikawa}
A. De Felice and S. Tsujikawa, {Living Rev. Relativity {\bf 13} (2010) 3.}

\bibitem{Khoury}
J. Khoury and A. Weltman, {Phys. Rev. {\bf D 69} (2004) 044026.}





\end{thebibliography}
\end{document}